\documentclass[preprint,1p,12pt]{elsarticle}
\usepackage[normalem]{ulem}
\usepackage{floatrow}

\usepackage{amssymb}
\usepackage{amsmath}

\usepackage{lineno}

\usepackage{booktabs}
\usepackage{threeparttable}
\usepackage{array}
\usepackage{upgreek}
\usepackage{bm}
\usepackage{xurl}

\journal{Journal of the Mechanics and Physics of Solids}
\biboptions{sort&compress}

\begin{document}
\begin{frontmatter}





\title{Mechanism of Band Gap Formation in Beam Networks}


\author[label1]{Kwangmin Lee} 
\author[label2]{Charles Emmett Maher} 
\author[label2]{Katherine A. Newhall} 
\author[label1,label3]{Ryan C. Hurley\corref{cor1}} 

\cortext[cor1]{Corresponding author}
\ead{rhurley6@jhu.edu}

\address[label1]{Department of Mechanical Engineering, Johns Hopkins University, Baltimore, Maryland 21218, USA}
\address[label2]{Department of Mathematics, University of North Carolina at Chapel Hill, Chapel Hill, North Carolina 27599, USA}
\address[label3]{Hopkins Extreme Materials Institute, Johns Hopkins University, Baltimore, Maryland 21218, USA}

\begin{abstract}
Band gaps are commonly attributed to Bragg scattering or local resonance, yet it remains unclear whether these mechanisms govern band gap formation in beam networks. In this work, we explain band gap formation in beam networks in terms of a new mechanism, geometry-induced coupling between deformation modes. Specifically, band gap onset arises from axial–bending coupling at lattice nodes and scales with the axial cutoff frequency of a one-dimensional periodic beam, whereas band gap termination is primarily governed by high-frequency rotational branches associated with beam geometry. This mechanism holds for both periodic and disordered beam networks. In periodic lattices, it manifests through beam orientations at lattice nodes, whereas in disordered networks it manifests through short-beam statistics arising from variations in beam length. Together, these results establish a unified mechanism for band gap formation across both periodic and disordered beam networks, providing new insight into the physical origin of band gaps in beam-network materials.
\end{abstract}

\begin{keyword}
Beam networks \sep
Phononic band gaps \sep
Band gap formation \sep
Geometry-induced coupling



\end{keyword}

\end{frontmatter}





\section{Introduction}\label{5sec:1}
There has been growing interest in beam-network materials for their exotic properties \cite{bertoldi2017,craster2023,torquato_multifunctional_2018,reid_auxetic_2018,rreid_ideal_2019,areyes-martinez_tuning_2022,mendels_systematic_2023,jiao2023,shen_autonomous_2024,maher2026}, including controlling wave propagation with phononic and photonic band gaps \cite{joannopoulos2008,florescu_designer_2009,man_photonic_2013,laude2015,froufe2016,siedentop_stealthy_2024}. For conventional materials, phononic band gaps are commonly attributed to Bragg scattering or local resonance \cite{Liu2012,hussein2014}. However, it remains unclear whether these same mechanisms explain band gap formation in beam-network materials. This question has become increasingly important as new techniques for manufacturing both ordered and disordered beam-network materials have emerged \cite{bertoldi2017,craster2023,moody2026}. Therefore, we seek a physical mechanism for band gap formation in such materials.

Bragg scattering is widely regarded as a primary mechanism for band gap formation in many periodic photonic and phononic materials \cite{joannopoulos1997,sigmund2003,lu2009,maldovan2013,maldovan2015}. Under Bragg scattering, waves reflected from periodically arranged material interfaces constructively interfere when the wavelength becomes comparable to the structural periodicity, producing strong reflection and suppressing wave propagation. As a result, the governing frequency scale is determined by the structural periodicity and the relevant wave speed. Even in disordered two-phase photonic systems, band gaps can arise from Bragg scattering when short-range order generates pronounced peaks in the structure factor, thereby establishing a characteristic length scale for band gap formation \cite{froufe2016}. However, additional long-range order may be required for such band gaps to persist in the thermodynamic limit \cite{klatt2022}.

Despite its success in explaining band gap formation in many photonic and phononic materials, the applicability of Bragg scattering to phononic beam networks remains unclear. In beam networks, waves propagate within a single constituent material and interact at lattice junctions, where beam geometry and orientation influence wave transmission, reflection, and mode conversion \cite{Svensson2010,Cremer2005,Leamy2012}. Accordingly, it remains unclear whether band gap formation in beam networks is fundamentally explained by Bragg scattering.

Another mechanism for band gap formation is local resonance, in which coupling between propagating waves and localized resonators suppresses wave propagation and produces subwavelength band gaps \cite{Liu2000,Liu2002,Liu2012}. This local resonance picture has been used to interpret band gap formation in periodic beam lattices \cite{phani2006,wang2015,warmuth2015}. In these studies, dispersion relations are often normalized by the flexural resonance frequency of an individual beam, implicitly treating the individual beam as the relevant resonant unit. This picture becomes less clear in higher-dimensional beam networks, where beams with different orientations are connected at lattice junctions. It is not straightforward what constitutes the relevant resonant unit for band gap formation. Moreover, even if an appropriate resonant unit is identified, it remains unclear whether band gap formation in beam networks is fundamentally explained by a local resonance mechanism.

While the dynamic behavior of a one-dimensional (1D) beam is well understood with analytical solutions under various boundary conditions \cite{rao2017}, wave propagation and band gap formation in higher-dimensional beam networks remain difficult to understand because they depend on lattice topology and beam geometry. Previous studies on two-dimensional (2D) and three-dimensional periodic beam lattices and related systems have shown that variations in these structural features can significantly affect band gap location and width \cite{phani2006,wang2015,warmuth2015,trainiti2016,chen2017}. However, despite extensive studies of higher-dimensional periodic beam lattices, a physical mechanism that explains band gap formation in periodic beam lattices and can be extended to disordered beam networks has not yet been established.

To investigate the physical mechanism governing band gap formation, a reliable formulation for computing dispersion relations is required. Bloch-periodic finite element (FE) formulations are widely used to compute dispersion relations in periodic beam lattices \cite{phani2006, sukumar2009}. However, the resulting dispersion relations can depend strongly on modeling assumptions, particularly the mass matrix formulation \cite{askes2024}. They may also depend on the choice of unit-cell representation, since different representations can introduce additional degrees of freedom (DOFs) that are not constrained by the Bloch phase relations, thereby altering the resulting dispersion relations. The influence of mass matrix formulation and unit-cell representation on the accuracy of dispersion relations has not been systematically examined, necessitating verification of FE-computed dispersion relations against analytical solutions such as those of a 1D periodic beam.

In this work, we show that band gap formation in 2D beam networks is governed by geometry-induced coupling between deformation modes. To establish this mechanism, we construct a Bloch-consistent FE formulation and verify its accuracy against the analytical dispersion relations of a 1D periodic Timoshenko beam. We then compute dispersion relations for representative 2D periodic beam lattices and show that band gap onset scales with the axial cutoff frequency of a 1D periodic beam due to geometry-induced axial–bending coupling at lattice nodes. We further show that variations in beam geometry primarily affect band gap termination associated with the rotational branch while leaving band gap onset largely unchanged. Finally, we extend the proposed mechanism to 2D disordered beam networks and show that band gap onset scales inversely with characteristic short-beam length.

The remainder of this paper is organized as follows. Section 2 presents the Bloch-periodic FE framework. Section 3 verifies the formulation against the analytical dispersion relations of a 1D periodic Timoshenko beam. Section 4 examines band gap formation and the underlying mechanism in 2D periodic beam networks. Section 5 investigates whether the proposed mechanism extends to disordered beam networks. Section 6 presents the conclusions.

\section{Bloch-Periodic Finite Element Framework}\label{5sec:2}
Wave propagation in periodic beam networks is analyzed using a Bloch-periodic FE framework. A single primitive unit cell is modeled with Timoshenko beam elements, and periodicity is enforced through Bloch phase relations between nodes connected by lattice translations. The resulting formulation leads to a generalized eigenvalue problem whose solutions provide the dispersion relations and, by sampling the first Brillouin zone, the density of states (DOS). The formulation is based on the standard Bloch FE framework for periodic lattices \cite{phani2006}, with modifications in the unit-cell representation and mass matrix formulation.

\begin{figure}[htbp]
\begin{center}
    \includegraphics[width=0.6\textwidth]{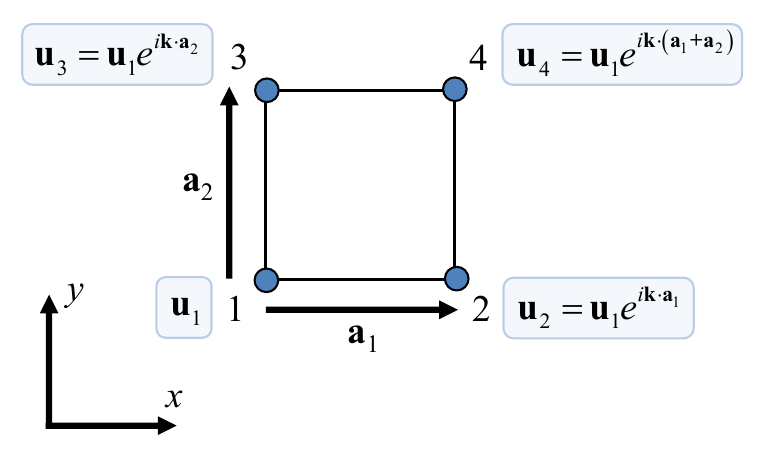}
    \caption[Generic two-dimensional unit cell with Bloch periodicity.]
    {Generic two-dimensional unit cell with Bloch periodicity. $\mathbf{u}_i$ represents the nodal displacement vector at node $i$, $\mathbf{a}_1$ and $\mathbf{a}_2$ denote the lattice vectors, and $\mathbf{k}$ denotes the wave vector.}
    \label{fig5:1}
\end{center}
\end{figure}

The periodic medium in two dimensions is constructed by repeating a primitive unit cell defined by lattice vectors $\mathbf{a}_1$ and $\mathbf{a}_2$, as illustrated in Fig.~\ref{fig5:1}. Each node carries three DOFs $(u, v, \theta)$ corresponding to $x$- and $y$-displacements, and rotation about the out-of-plane axis. The nodal displacement vector at node $i$ is defined as $\mathbf{u}_i = (u_i, v_i, \theta_i)$. Beams are represented by single Timoshenko elements connecting primitive unit cell nodes. At each node, connected beams share both translational and rotational DOFs, corresponding to a rigid joint. The same formulation also applies to 1D periodic beams by restricting the lattice to a single translation vector.

Figure~\ref{fig5:1} illustrates a generic 2D unit cell and the associated Bloch phase relations applied to boundary nodes. For a wave vector $\mathbf{k}$, nodes connected by lattice translations are not independent. For example, if node 2 is obtained from node 1 by translation $\mathbf{a}_1$, their nodal displacement vectors satisfy $\mathbf{u}_2 = \mathbf{u}_1 e^{i \mathbf{k} \cdot \mathbf{a}_1}$, with analogous relations for translations by $\mathbf{a}_2$ and $\mathbf{a}_1 + \mathbf{a}_2$. These relations enforce Bloch periodicity across opposite boundaries of the unit cell. 

Accordingly, the unit cell displacement vector $\mathbf{q}$, collecting all nodal DOFs $(u, v, \theta)$, can be expressed in terms of a reduced set of independent DOFs through a $\mathbf{k}$-dependent transformation matrix $\mathbf{T}(\mathbf{k})$, such that
\begin{equation}\label{eqn:5_1}
\mathbf{q} = \mathbf{T}(\mathbf{k})\tilde{\mathbf{q}}.
\end{equation}
Here, $\tilde{\mathbf{q}}$ denotes the reduced displacement vector associated with the independent DOFs, obtained by eliminating dependent DOFs through Bloch periodicity relations.

Element stiffness matrices $\mathbf{K}_\text{e}$ are constructed using the standard Timoshenko beam formulation, incorporating axial deformation and shear-corrected bending behavior \cite{przemieniecki1985}, and element mass matrices $\mathbf{M}_\text{e}$ are constructed using the hybrid mass matrix formulation introduced in Section~\ref{5sec:3}. After assembly, the unit-cell stiffness and mass matrices are denoted by $\mathbf{K}$ and $\mathbf{M}$. The equations of motion for the unreduced unit-cell system read
\begin{equation}\label{eqn:5_2}
\mathbf{M}\ddot{\mathbf{q}} + \mathbf{K}\mathbf{q} = 0,
\end{equation}
where overdots denote time derivatives.

Substituting $\mathbf{q} = \mathbf{T}\tilde{\mathbf{q}}$ in Eq.~\eqref{eqn:5_2} and premultiplying by $\mathbf{T}^{\mathrm{H}}(\mathbf{k})$ yields the Bloch-reduced system
\begin{equation}\label{eqn:5_3}
\mathbf{M}_{\mathrm{red}}\,\ddot{\tilde{\mathbf{q}}}
+ \mathbf{K}_{\mathrm{red}}\,\tilde{\mathbf{q}} = 0
\end{equation}
with
\begin{equation}\label{eqn:5_4}
\mathbf{K}_{\mathrm{red}} = \mathbf{T}^{\mathrm{H}}(\mathbf{k})\,\mathbf{K}\,\mathbf{T}(\mathbf{k}), \qquad
\mathbf{M}_{\mathrm{red}} = \mathbf{T}^{\mathrm{H}}(\mathbf{k})\,\mathbf{M}\,\mathbf{T}(\mathbf{k}),
\end{equation}
where $(\cdot)^{\mathrm{H}}$ denotes the Hermitian (complex conjugate) transpose.

Assuming time-harmonic motion of the form $\tilde{\mathbf{q}} = \bm{\uppsi} e^{i\omega t}$, the problem reduces to the generalized eigenvalue problem
\begin{equation}\label{eqn:5_5}
\mathbf{K}_{\mathrm{red}}(\mathbf{k})\,\bm{\uppsi}
= \omega^{2}(\mathbf{k})\,\mathbf{M}_{\mathrm{red}}(\mathbf{k})\,\bm{\uppsi},
\end{equation}
where $\omega(\mathbf{k})$ denotes the angular eigenfrequency and $\bm{\uppsi}$ denotes the corresponding eigenvector.

\section{One-dimensional Analytical Dispersion and Finite Element Verification}\label{5sec:3}
To assess the Bloch-periodic FE formulation introduced in Section~\ref{5sec:2}, we compare the FE dispersion relations with the closed-form analytical solution of a 1D periodic Timoshenko beam. The comparison identifies the appropriate primitive unit cell representation and mass matrix formulation prior to extending the framework to 2D beam networks.

\subsection{Analytical dispersion relations}\label{5sec:3_1}
\begin{figure}[htbp]
\begin{center}
    \includegraphics[width=1.0\textwidth]{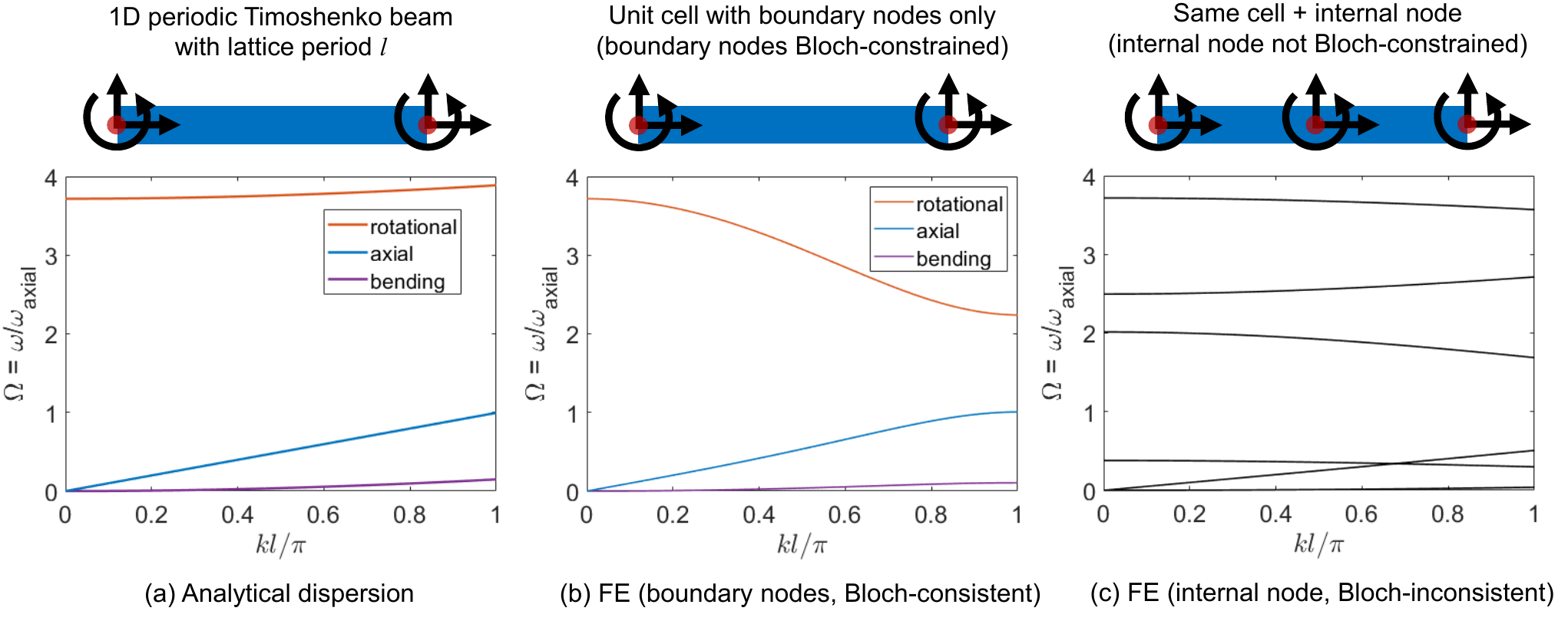}
    \caption[Unit-cell representations and dispersion relations for a one-dimensional periodic beam.]
    {Unit-cell schematics and dispersion relations for a 1D periodic Timoshenko beam: (a) analytical, (b) Bloch-consistent FE, and (c) Bloch-inconsistent FE. Frequencies are normalized by the axial cutoff frequency $\omega_{\mathrm{axial}}$.}
    \label{fig5:2}
\end{center}
\end{figure}

We consider an infinite 1D periodic Timoshenko beam with lattice period $l$, illustrated in Fig.~\ref{fig5:2}(a). Timoshenko beam theory accounts for transverse shear deformation and rotary inertia and is applicable to beams of moderate slenderness where these effects are non-negligible \cite{rao2017}. In this setting, the primitive unit cell consists of a single beam segment of length $l$. For each wavenumber $k$, three dispersion branches exist: one axial, one bending, and one rotational.
The axial displacement $u(x, t)$, where $x$ is the coordinate along the beam axis and $t$ denotes time, satisfies \cite{rao2017}
\begin{equation}\label{eqn:5_6}
\rho A\,\ddot{u} - E A\,u'' = 0,
\end{equation}
where $E$ is Young’s modulus, $\rho$ is the material density, $A$ is the cross-sectional area, and primes denote derivatives with respect to $x$. We assume a harmonic solution of the form 
\begin{equation}\label{eqn:5_7}
u(x,t) = U\,e^{i(kx - \omega t)},
\end{equation}
where $U$ is a constant amplitude. Substituting Eq.~\eqref{eqn:5_7} into Eq.~\eqref{eqn:5_6} yields
\begin{equation}\label{eqn:5_8}
\omega(k) = |k|\sqrt{\frac{E}{\rho}}.
\end{equation}
Bloch periodicity restricts $k$ to the first Brillouin zone, $k \in [-\pi/l, \pi/l]$. The axial branch reaches its maximum at the zone boundary $k = \pm \pi/l$, defining the axial cutoff frequency
\begin{equation}\label{eqn:5_9}
\omega_{\mathrm{axial}} = \frac{\pi}{l}\sqrt{\frac{E}{\rho}}.
\end{equation}
The transverse displacement $v(x, t)$ and the rotation $\theta(x, t)$ about the out-of-plane axis are governed by the coupled Timoshenko beam equations \cite{rao2017}, which can be combined into a higher-order equation accounting for shear deformation and rotary inertia:
\begin{equation}\label{eqn:5_10}
EI\,v'''' + \rho A\,\ddot{v}
- \rho I\left(1 + \frac{E}{\kappa G}\right)\ddot{v}''
+ \frac{\rho^{2} I}{\kappa G}\,\overset{....}{v} = 0,
\end{equation}
where $I$ is the second moment of area, $G$ is the shear modulus, and $\kappa$ is the shear correction factor \cite{cowper1966}. We assume a harmonic solution of the form
\begin{equation}\label{eqn:5_11}
v(x,t) = V\,e^{i(kx - \omega t)},
\end{equation}
where $V$ is a constant amplitude, and we define $s = \omega^2$. Substituting Eq.~\eqref{eqn:5_11} into Eq.~\eqref{eqn:5_10} gives
\begin{equation}\label{eqn:5_12}
\frac{\rho^{2} I}{\kappa G}\,s^{2}
- \left[\rho A + \rho I\left(1 + \frac{E}{\kappa G}\right)k^{2}\right] s
+ EI\,k^{4} = 0,
\end{equation}
which yields two positive solutions
\begin{equation}\label{eqn:5_13}
s_{\pm}(k) =
\frac{
\left[ A + I\left(1 + \frac{E}{\kappa G}\right)k^{2} \right]
\pm
\sqrt{
\left[ A + I\left(1 + \frac{E}{\kappa G}\right)k^{2} \right]^{2}
- 4\,\frac{I^{2}E}{\kappa G}\,k^{4}
}
}{
2\rho I/(\kappa G)
}.
\end{equation}
The corresponding eigenfrequencies are
\begin{equation}\label{eqn:5_14}
\omega_{\pm}(k) = \sqrt{s_{\pm}(k)}.
\end{equation}
The lower branch $\omega_{-}(k)$ corresponds to bending, and the upper branch $\omega_{+}(k)$ corresponds to rotation. The bending and rotational branches reach their maxima at the Brillouin-zone boundary $k = \pm \pi/l$, defining the corresponding cutoff frequencies:
\begin{equation}\label{eqn:5_15}
\omega_{\mathrm{bend}} = \omega_{-}(k = \pm \pi/l), \qquad
\omega_{\mathrm{rot}} = \omega_{+}(k = \pm \pi/l).
\end{equation}
One notable difference between the axial branch and the bending and rotational branches is the dependence of their cutoff frequencies on beam geometry. Unlike the axial cutoff frequency $\omega_{\mathrm{axial}}$, which depends only on beam length and material properties (Eq.~\eqref{eqn:5_9}), the cutoff frequencies $\omega_{\mathrm{bend}}$ and $\omega_{\mathrm{rot}}$ depend on cross-sectional geometry, as the geometric terms $A$ and $I$ remain in Eqs.~\eqref{eqn:5_13}--\eqref{eqn:5_15}.

\subsection{Bloch-periodic finite element formulation for a one-dimensional unit cell}\label{5sec:3_2}
A primitive 1D unit cell, illustrated in Fig.~\ref{fig5:2}(b), consists of a single Timoshenko beam element of length $l$ connecting two boundary nodes (left and right). Each node carries three DOFs $(u, v, \theta)$, corresponding to axial displacement, transverse displacement, and rotation about the out-of-plane axis. Let $\mathbf{u}_L$ and $\mathbf{u}_R$ denote the nodal displacement vectors at the left and right boundaries of the unit cell. For a Bloch wave with wavenumber $k$, the boundary displacement vectors satisfy
\begin{equation}\label{eqn:5_16}
\mathbf{u}_R = \mathbf{u}_L\,e^{ikl}.
\end{equation}
Enforcing this relation yields the $k$-dependent reduction given in Eq.~\eqref{eqn:5_1}.

To assess the effect of introducing an internal node, we also consider a unit cell containing a central internal node whose DOFs are not subject to Bloch phase constraints (Fig.~\ref{fig5:2}(c)). This configuration is obtained by introducing an internal node that subdivides the beam into two element segments. At this internal node, the nodal DOFs are shared by adjacent element segments, corresponding to a rigid connection.

\subsection{Hybrid mass matrix}\label{5sec:3_3}
The influence of mass matrix choice on eigenfrequencies in FE models is well known \cite{kim1993, chandrupatla2012fem}. Since dispersion relations are obtained from these eigenfrequencies, they likewise depend on the choice of mass matrix. A quantitative comparison of analytical dispersion relations and those obtained from the FE formulation using conventional consistent and lumped mass matrices is provided in~\ref{sec:appendix-D}. The results show that the consistent mass matrix alters the relative branch ordering over part of the Brillouin zone relative to the ordering in the analytical solution, whereas the lumped mass matrix preserves the branch ordering present in the analytical solution but systematically underestimates the frequencies of all branches. To preserve the branch ordering and relative frequency scaling consistent with the analytical dispersion relations, we formulate a hybrid mass matrix.

For the axial contribution, we construct the element-level axial mass block $\mathbf{M}_\text{ax}$ as a linear combination of the consistent mass matrix $\mathbf{M}_\text{C}$ and the lumped mass matrix $\mathbf{M}_\text{L}$, following the linear-combination strategy discussed by \cite{kim1993}:
\begin{equation}\label{eqn:5_17}
\mathbf{M}_{\mathrm{ax}} = \alpha \mathbf{M}_{\mathrm{C}} + \beta \mathbf{M}_{\mathrm{L}}
= \frac{\alpha \rho A l}{6}
\begin{bmatrix}
2 & 1 \\
1 & 2
\end{bmatrix}
+ \frac{\beta \rho A l}{2}
\begin{bmatrix}
1 & 0 \\
0 & 1
\end{bmatrix},
\end{equation}
where $\alpha + \beta = 1$ ensures mass conservation. Here, we set $\alpha = 1/2$ and $\beta = 1/2$ so that the FE axial cutoff frequency closely matches the analytical axial cutoff frequency $\omega_{\mathrm{axial}}$. The resulting FE axial cutoff frequency is not highly sensitive to the choice of $\alpha$ and $\beta$ within a reasonable range, as the selected values balance the overestimation from the consistent mass matrix and the underestimation from the lumped mass matrix.

For the transverse and rotational components, the element-level mass block $\mathbf{M}_\text{tr}$ is constructed from the classical lumped mass matrix \cite{askes2024}:
\begin{equation}\label{eqn:5_18}
\mathbf{M}_{\mathrm{tr}} =
\frac{\rho l}{2}
\begin{bmatrix}
A & 0 & 0 & 0 \\
0 & I & 0 & 0 \\
0 & 0 & A & 0 \\
0 & 0 & 0 & I
\end{bmatrix}.
\end{equation}
However, this formulation underestimates the rotational branch. This discrepancy can be attributed to the lack of dynamic moment equilibrium in the classical lumped mass matrix formulation for Timoshenko beams \cite{laier2007}. To address this discrepancy, we adopt the effective rotational inertia proposed by \cite{laier2007}. Unlike the original formulation, which applies the correction to both the mass and rotational inertia, we apply it only to the rotational inertia term, since correcting the mass terms in Eq.~\eqref{eqn:5_18} overestimates the bending branch. The transverse and rotational mass block $\mathbf{M}_{\mathrm{tr}}$ then takes the form:
\begin{equation}\label{eqn:5_19}
\mathbf{M}_{\mathrm{tr}} =
\frac{\rho l}{2}
\begin{bmatrix}
A & 0 & 0 & 0 \\
0 & \dfrac{\Phi}{1 + \Phi}\, I & 0 & 0 \\
0 & 0 & A & 0 \\
0 & 0 & 0 & \dfrac{\Phi}{1 + \Phi}\, I
\end{bmatrix},
\end{equation}
where $\Phi = 12EI/(G \kappa A l^2)$. 

The element-level mass matrix $\mathbf{M}_\text{e}$ is assembled by placing the axial mass block $\mathbf{M}_\text{ax}$ in the DOF positions (1, 4) and the transverse mass block $\mathbf{M}_\text{tr}$ in the DOF positions (2, 3, 5, 6), consistent with the nodal ordering $(u_1, v_1, \theta_1, u_2, v_2, \theta_2)$.

\subsection{Verification against the analytical dispersion relations}\label{5sec:3_4}
To verify the Bloch-periodic FE formulation against the analytical dispersion relations derived in Section~\ref{5sec:3_1}, we consider a Timoshenko beam with length $l = 25\,\mu\mathrm{m}$ and circular cross section of radius $r = 2.5\,\mu\mathrm{m}$. The Young’s modulus is $E = 200\,\mathrm{GPa}$, Poisson’s ratio is $\nu = 0.3$, density is $\rho = 8000\,\mathrm{kg/m^3}$, and the shear modulus is $G = E/(2(1+\nu))$. For a circular cross-section, the shear correction factor is taken as $\kappa = 6(1+\nu)/(1+6\nu)$, following \cite{cowper1966}. The geometric and material parameters are chosen as representative values, and the resulting dispersion relations exhibit the same qualitative behavior for other parameter sets. The analytical and Bloch-periodic FE dispersion relations are compared in Fig.~\ref{fig5:2}, where the angular frequency is normalized by the axial cutoff frequency $\omega_{\mathrm{axial}}$ defined in Eq.~\eqref{eqn:5_9}, i.e., $\Omega = \omega / \omega_{\mathrm{axial}}$.

Figure~\ref{fig5:2}(a) presents the analytical dispersion relations. Figure~\ref{fig5:2}(b) shows the FE results obtained using a primitive unit cell composed of a single Timoshenko beam element with Bloch constraints imposed at the boundary nodes. In this configuration, the FE formulation produces three dispersion branches for each wavenumber $k$, consistent with the analytical solution: one axial, one bending, and one rotational. The axial and bending branches show close agreement with the analytical branches across the Brillouin zone. However, the rotational branch exhibits a noticeable deviation at higher frequencies that is not eliminated by the hybrid mass matrix. Relative to both classical lumped and consistent mass matrix formulations (see~\ref{sec:appendix-D}, Fig.~\ref{figD:1}), the present hybrid mass matrix formulation provides a rotational branch that more closely matches the analytical solution, primarily due to the rotary inertia correction proposed by \cite{laier2007}. Nevertheless, a residual deviation persists due to the discrete mass representation inherent to the FE formulation. This deviation is confined to the high-frequency regime and does not affect the band gap onset frequency of interest. However, it may influence the quantitative prediction of the band gap termination frequency and, consequently, the band gap width.

Figure~\ref{fig5:2}(c) shows the FE results for a unit cell that includes an additional internal node at the beam center, whose DOFs are not constrained by the Bloch phase relation. In this case, six independent DOFs exist per unit cell, leading to six dispersion branches at each $k$. The number of branches therefore no longer matches the three analytical branches of the 1D periodic Timoshenko beam.

To understand this discrepancy, we examine the Bloch form of the displacement field. The axial displacement field takes the form given in Eq.~\eqref{eqn:5_7}:
\[
u(x,t) = U\,e^{i(kx - \omega t)}.
\]
This corresponds to a Bloch form $u(x) = p(x)e^{ikx}$, where the periodic component $p(x)$ is spatially constant within the primitive cell. Hence, the spatial dependence inside the cell is entirely described by the phase factor $e^{ikx}$. Consequently, the displacement at any interior point is uniquely determined by the boundary values through the relation $u(x_b + \Delta x) = u(x_b)e^{ik\Delta x}$ for any $\Delta x$ within the cell, where $x_b$ denotes a boundary point of the primitive cell. The same reasoning applies to the transverse displacement and rotational fields, which admit the same Bloch form.

Introducing an internal node whose DOFs are not constrained by the Bloch phase relation permits additional spatial variation inside the unit cell that is not uniquely determined by the boundary values through the Bloch relation. The resulting extra dispersion branches in Fig.~\ref{fig5:2}(c) therefore do not correspond to admissible Bloch modes of the primitive periodic beam, but instead arise from these unconstrained internal-node DOFs. Accordingly, a primitive unit cell consisting of a single beam element with Bloch constraints imposed only at the boundary nodes provides the physically consistent discretization for the 1D periodic beam. Based on this observation, the minimal primitive unit cell representation together with the hybrid mass matrix is adopted for the subsequent 2D band gap formation analysis.

\section{Band Gap Formation in Two-dimensional Periodic Beam Networks}\label{5sec:4}
Section~\ref{5sec:3} verified that the proposed FE formulation reproduces the analytical dispersion relations of a 1D periodic Timoshenko beam. We now apply this framework to 2D periodic beam networks and analyze the dispersion relations of square, triangular, honeycomb, and Kagome beam lattices to investigate the mechanism underlying the formation of the first complete band gap. We also examine the influence of cross-sectional geometry by varying the beam slenderness ratio.

\subsection{Primitive unit cells}\label{5sec:4_1}
\begin{figure}[htbp]
\begin{center}
    \includegraphics[width=1.0\textwidth]{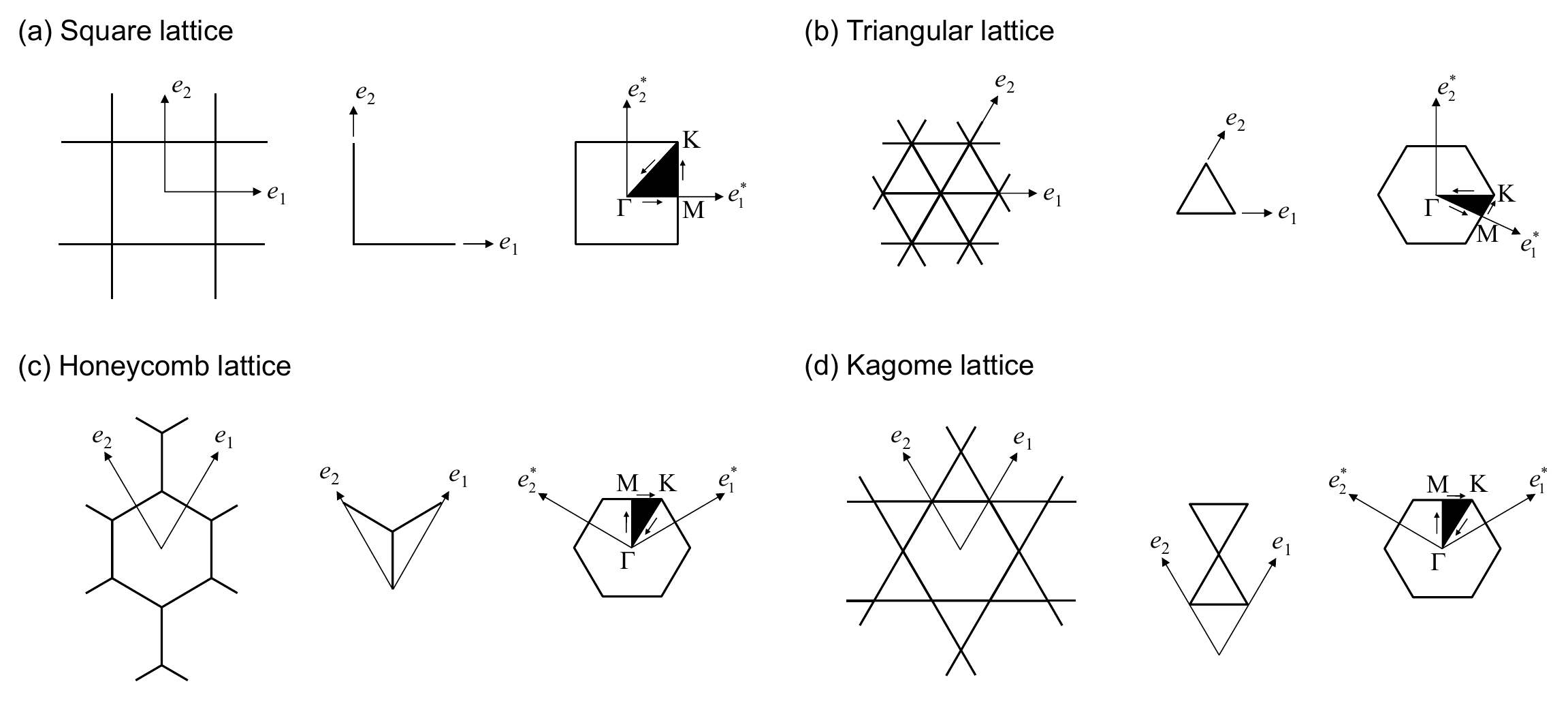}
    \caption[Periodic beam lattices and reciprocal representations.]
    {Periodic beam lattices and their reciprocal representations: for each lattice, the left panel shows the infinite lattice, the middle panel the selected primitive unit cell, and the right panel the corresponding first Brillouin zone in reciprocal space. The basis vectors of the direct lattice $(\mathbf{e}_1, \mathbf{e}_2)$ and reciprocal lattice $(\mathbf{e}_1^*, \mathbf{e}_2^*)$ are indicated as defined in Table~\ref{tab:primitive_vectors}. The high-symmetry points $\Gamma$, M, and K are defined along the high-symmetry path used for the dispersion calculations.}
    \label{fig5:3}
\end{center}
\end{figure}

Figure~\ref{fig5:3} presents the four periodic beam lattices considered in this study: square, triangular, honeycomb, and Kagome lattices. For each lattice, the left panel shows the infinite lattice, the middle panel shows the selected primitive unit cell, and the right panel shows the corresponding first Brillouin zone in reciprocal space. The basis vectors of the direct lattice $(\mathbf{e}_1, \mathbf{e}_2)$ and reciprocal lattice $(\mathbf{e}_1^*, \mathbf{e}_2^*)$ are indicated, as defined in Table~\ref{tab:primitive_vectors}, and the high-symmetry points $\Gamma$, M, and K are indicated in the Brillouin zone.

The lattices and their reciprocal representations are consistent with those in \cite{phani2006}. However, the primitive unit cells selected here differ in one essential aspect. In the unit-cell construction of Phani et al. \cite{phani2006}, the unit-cell boundaries pass through the interior of certain beams, such that FE modeling introduces internal nodes with DOFs not constrained by the Bloch phase relation. In contrast, the primitive unit cells in the present work are constructed such that each beam is represented by a single Timoshenko element connecting primitive lattice nodes, thereby avoiding internal nodes whose DOFs are not constrained by the Bloch phase relation. This definition is motivated by the verification results in Section~\ref{5sec:3}, where it was shown that introducing internal DOFs not constrained by the Bloch phase relation leads to spurious modes in the dispersion relations.

\begin{table}[htbp]
\centering
\begin{threeparttable}
\setlength{\tabcolsep}{10pt}
\renewcommand{\arraystretch}{1.3}

\begin{tabular}{>{\centering\arraybackslash}m{0.18\textwidth}
                >{\centering\arraybackslash}m{0.36\textwidth}
                >{\centering\arraybackslash}m{0.36\textwidth}}
\toprule \toprule
Topology & Direct lattice & Reciprocal lattice \\[2pt] 
\toprule \toprule

Square lattice
&
\begin{tabular}{@{}c@{}}
$\mathbf{e}_1 = l\,\mathbf{i}$ \\[3pt]
$\mathbf{e}_2 = l\,\mathbf{j}$
\end{tabular}
&
\begin{tabular}{@{}c@{}}
$\mathbf{e}_1^{*} = \frac{2\pi}{l}\,\mathbf{i}$ \\[3pt]
$\mathbf{e}_2^{*} = \frac{2\pi}{l}\,\mathbf{j}$
\end{tabular}
\\

Triangular lattice
&
\begin{tabular}{@{}c@{}}
$\mathbf{e}_1 = l\,\mathbf{i}$ \\[3pt]
$\mathbf{e}_2 = l\left(\frac{1}{2}\mathbf{i}+\frac{\sqrt{3}}{2}\mathbf{j}\right)$
\end{tabular}
&
\begin{tabular}{@{}c@{}}
$\mathbf{e}_1^{*} = \frac{2\pi}{l}\left(\mathbf{i}-\frac{1}{\sqrt{3}}\mathbf{j}\right)$ \\[3pt]
$\mathbf{e}_2^{*} = \frac{2\pi}{l}\left(\frac{2}{\sqrt{3}}\mathbf{j}\right)$
\end{tabular}
\\

Honeycomb lattice
&
\begin{tabular}{@{}c@{}}
$\mathbf{e}_1 = \sqrt{3}l\left(\frac{1}{2}\mathbf{i}+\frac{\sqrt{3}}{2}\mathbf{j}\right)$ \\[3pt]
$\mathbf{e}_2 = \sqrt{3}l\left(-\frac{1}{2}\mathbf{i}+\frac{\sqrt{3}}{2}\mathbf{j}\right)$
\end{tabular}
&
\begin{tabular}{@{}c@{}}
$\mathbf{e}_1^{*} = \frac{2\pi}{\sqrt{3}l}\left(\mathbf{i}+\frac{1}{\sqrt{3}}\mathbf{j}\right)$ \\[3pt]
$\mathbf{e}_2^{*} = \frac{2\pi}{\sqrt{3}l}\left(-\mathbf{i}+\frac{1}{\sqrt{3}}\mathbf{j}\right)$
\end{tabular}
\\

Kagome lattice
&
\begin{tabular}{@{}c@{}}
$\mathbf{e}_1 = 2l\left(\frac{1}{2}\mathbf{i}+\frac{\sqrt{3}}{2}\mathbf{j}\right)$ \\[3pt]
$\mathbf{e}_2 = 2l\left(-\frac{1}{2}\mathbf{i}+\frac{\sqrt{3}}{2}\mathbf{j}\right)$
\end{tabular}
&
\begin{tabular}{@{}c@{}}
$\mathbf{e}_1^{*} = \frac{\pi}{l}\left(\mathbf{i}+\frac{1}{\sqrt{3}}\mathbf{j}\right)$ \\[3pt]
$\mathbf{e}_2^{*} = \frac{\pi}{l}\left(-\mathbf{i}+\frac{1}{\sqrt{3}}\mathbf{j}\right)$
\end{tabular}
\\

\bottomrule
\end{tabular}

\caption[Primitive translation vectors of the four lattices]
{Primitive translation vectors of the four lattices. $l$ denotes the length of each beam of the lattice; $\mathbf{i}$ and $\mathbf{j}$ are the Cartesian unit vectors in the $x$--$y$ plane. Note that $\mathbf{e}_i \cdot \mathbf{e}_j^{*} = 2\pi \delta_{ij}$, where $\delta_{ij}$ is the Kronecker delta.}
\label{tab:primitive_vectors}

\end{threeparttable}
\end{table}

\subsection{Band gap onset and axial--bending coupling}\label{5sec:4_2}
The dispersion relations of the 2D periodic beam networks are computed using the FE formulation developed in Sections~\ref{5sec:2}--\ref{5sec:3} together with the material properties and beam cross-sectional parameters used in Section~\ref{5sec:3_4}. For each prescribed wavevector along the high-symmetry k-path $\Gamma$–M–K–$\Gamma$ in the first Brillouin zone shown in Fig.~\ref{fig5:3}, the resulting eigenvalue problem is solved to obtain the angular eigenfrequencies. Repeating this procedure along the entire k-path yields the dispersion relations for each lattice.

\begin{figure}[htbp]
\begin{center}
    \includegraphics[width=0.9\textwidth]{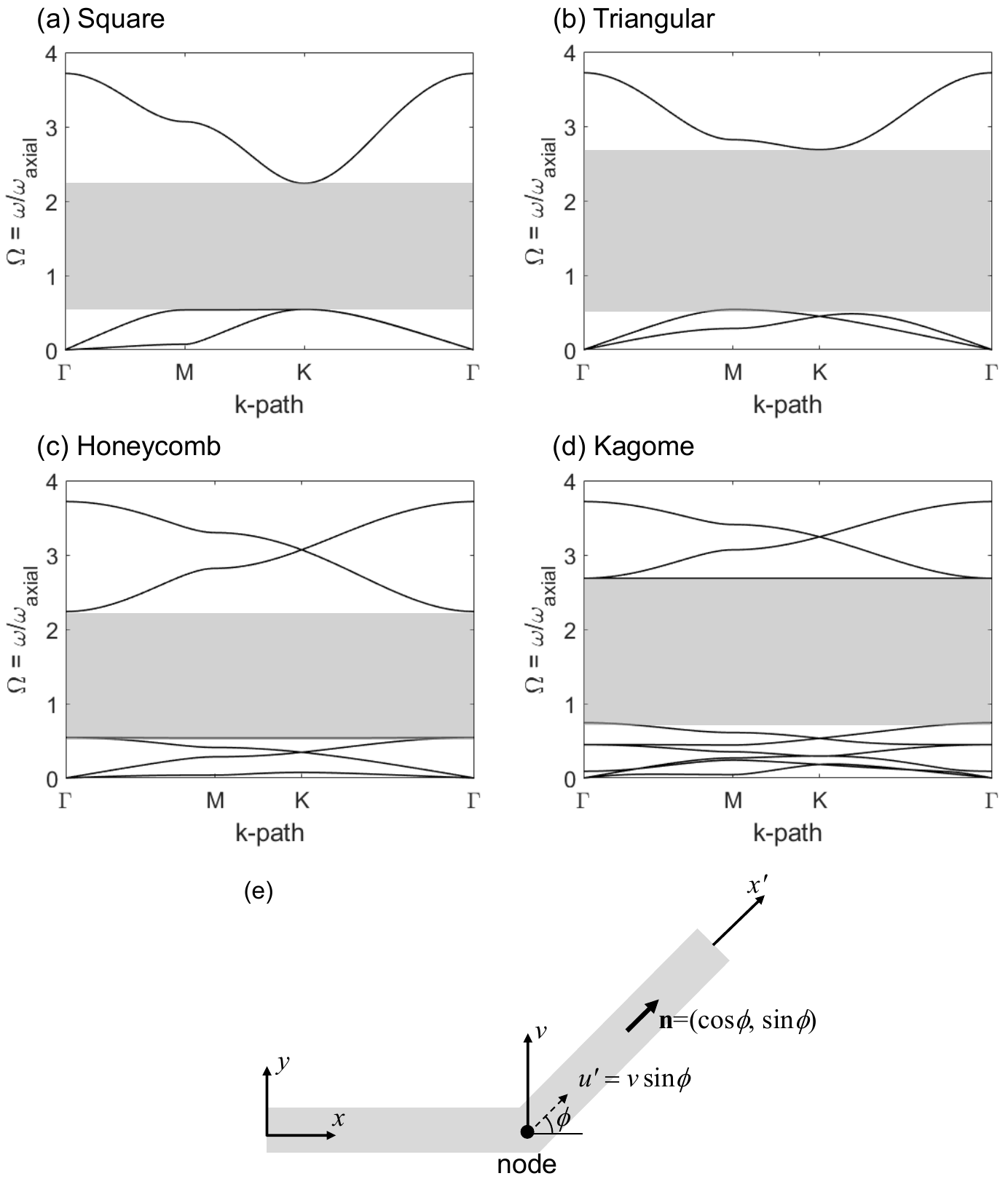}
    \caption[Dispersion relations and axial--bending coupling in two-dimensional beam lattices.]
    {Dispersion relations of 2D periodic beam lattices: (a) square lattice, (b) triangular lattice, (c) honeycomb lattice, and (d) Kagome lattice. Frequencies are normalized by the axial cutoff frequency $\omega_{\mathrm{axial}}$ of a 1D periodic beam, defined in Eq.~\eqref{eqn:5_9}, and plotted along the high-symmetry k-path $\Gamma$–M–K–$\Gamma$ in the first Brillouin zone shown in Fig.~\ref{fig5:3}. The shaded gray regions indicate the first complete band gaps. (e) Schematic illustrating the geometric origin of coupling between deformation modes at a lattice node. A transverse displacement $v$ of the horizontal beam induces an axial deformation $u' = v \sin\phi$ in an inclined beam. The unit vector $\mathbf{n}$ denotes the beam-axis direction.}
    \label{fig5:4}
\end{center}
\end{figure}

Figure~\ref{fig5:4}(a)--(d) presents the dispersion relations of the square, triangular, honeycomb, and Kagome lattices. For the square, triangular, and honeycomb lattices, the first complete band gap onset occurs at approximately $\Omega = \omega / \omega_{\mathrm{axial}} \approx 0.5$, whereas for the Kagome lattice the onset shifts to approximately $\Omega \approx 0.75$, where $\omega_{\mathrm{axial}}$ is the axial cutoff frequency of a 1D periodic beam of length $l$ defined in Eq.~\eqref{eqn:5_9}. 

Under a Bragg scattering picture, band gap formation is expected when reflected waves constructively interfere as the wavelength $\lambda$ becomes comparable to the structural periodicity $(\lambda \sim 2l)$, giving a characteristic Bragg frequency $\omega_{\mathrm{Bragg}} \sim c_{\mathrm{axial}}\pi/l \,(=\omega_{\mathrm{axial}})$, where $c_{\mathrm{axial}}=\sqrt{E/\rho}$ is the axial wave speed and $l$ is the beam length. However, the observed band gap onset occurs at frequencies lower than $\omega_{\mathrm{Bragg}}$, specifically near $0.5\omega_{\mathrm{axial}}$ for the square, triangular, and honeycomb lattices, and near $0.75\omega_{\mathrm{axial}}$ for the Kagome lattice. Furthermore, the onset frequency varies across different beam lattice topologies, indicating that band gap onset depends on lattice topology rather than only the beam length and axial wave speed. This dependence on lattice topology is inconsistent with a Bragg scattering picture, in which the Bragg frequency is determined primarily by the beam length and axial wave speed. These observations indicate that band gap onset cannot be explained solely by Bragg scattering.

Under a local resonance picture, localized resonances are typically associated with nearly flat dispersion branches near band gaps \cite{wang2015}. However, except for the honeycomb lattice, no pronounced flat dispersion branch is observed near the band gap onset frequency in the dispersion relations computed using the present primitive unit cell. This behavior is inconsistent with a conventional local resonance picture. Furthermore, if band gap formation were governed by local resonance of individual beams, the onset frequency would be expected to follow the fixed-fixed flexural resonance frequency of a beam, approximately \(22.4\sqrt{EI/(m l^4)}\). This frequency scale is comparable to \(\omega_{\mathrm{bend}}\). However, the observed onset frequencies instead scale with \(\omega_{\mathrm{axial}}\).

This observed scaling with $\omega_{\mathrm{axial}}$ instead arises from geometry-induced coupling between deformation modes at lattice nodes. The geometric origin of this coupling is illustrated in Fig.~\ref{fig5:4}(e), where a transverse displacement $v$ of the horizontal beam induces an axial displacement in an inclined beam according to $u' = v \sin\phi$. Although the imposed motion is purely transverse in the horizontal beam, it generates an axial strain component along the beam axis in the inclined beam.

To formalize this mechanism, we now consider the general case of multiple beams connected to a node. For each beam $i$ of length $l_i$, the axial stiffness scales as $K_{i,\mathrm{axial}} \sim EA / l_i$, and the bending stiffness scales as $K_{i,\mathrm{bend}} \sim EI / l_i^3$. Let $\mathbf{n}_i = (\cos\phi_i, \sin\phi_i)$ denote the unit vector along the beam axis and $\mathbf{t}_i$ the in-plane unit vector orthogonal to it. The translational stiffness tensor of beam $i$ in a global coordinate frame can then be expressed as
\begin{equation}\label{eqn:5_20}
\mathbf{K}_{\mathrm{tr},i} \approx K_{i,\mathrm{axial}}\left( \mathbf{n}_i \otimes \mathbf{n}_i \right) + K_{i,\mathrm{bend}}\left( \mathbf{t}_i \otimes \mathbf{t}_i \right).
\end{equation}
Summing over all beams connected to the node gives an effective translational stiffness tensor
\begin{equation}\label{eqn:5_21}
\mathbf{K}_{\mathrm{tr,eff}} = \sum\limits_i \mathbf{K}_{\mathrm{tr},i}.
\end{equation}
Projecting onto the $y$-direction gives
\begin{equation}\label{eqn:5_22}
K_{{\rm{eff}}}^{\left( y \right)}
=
\mathbf{e}_y^{\rm{T}}
\mathbf{K}_{{\rm{tr,eff}}}
\mathbf{e}_y
=
\sum\limits_i
\mathbf{e}_y^{\rm{T}}
\mathbf{K}_{{\rm{tr,}}i}
\mathbf{e}_y
\approx
\sum\limits_i
\left(
K_{i,{\rm{axial}}}{{\sin }^2}{\phi _i}
+
K_{i,{\rm{bend}}}{{\cos }^2}{\phi _i}
\right),
\end{equation}
where $\mathbf{e}_y = (0,1)$. Since $K_{i,\mathrm{axial}} \gg K_{i,\mathrm{bend}}$, the effective stiffness scales as $\sum_i K_{i,\mathrm{axial}} \sin^2 \phi_i$, indicating that axial contributions dominate and are weighted by beam orientations. 

For periodic lattices, the influence of beam orientations is quantified by averaging $\sin^2 \phi_i$ over all beams connected to a node, defining the nodal orientation factor as $\frac{1}{N_b} \sum_i \sin^2 \phi_i$ where $N_b$ denotes the number of beams connected to the node. For the square, triangular, and honeycomb lattices, this factor equals 0.5, consistent with the normalized band gap onset observed in Fig.~\ref{fig5:4}(a)--(c) at $\Omega \approx 0.5$. For the Kagome lattice, inequivalent nodal configurations exist within the unit cell: the central node yields a nodal orientation factor of 0.75, whereas boundary nodes yield $3/8$. The larger orientation factor leads to greater effective stiffness and corresponds to the band gap onset at $\Omega \approx 0.75$ in Fig.~\ref{fig5:4}(d).

This interpretation differs from that of a 1D periodic beam. As shown in the analytical dispersion relation of the 1D periodic beam (Fig.~\ref{fig5:2}(a)), the bending branch terminates at $\omega_{\mathrm{bend}}$. Above this frequency, only the axial branch remains, whose substantially larger dispersion slope gives rise to a markedly lower density of states. As a result, pseudogap-like behavior emerges above $\omega_{\mathrm{bend}}$ and persists until the complete band gap forms at $\omega_{\mathrm{axial}}$. In contrast, geometry-induced axial--bending coupling in 2D beam networks causes axial stiffness to dominate the effective network response near band gap onset. Consequently, bending-based interpretations derived from 1D periodic beams do not directly extend to higher-dimensional beam networks.

\subsection{Effect of beam slenderness ratio}\label{5sec:4_3}
\begin{figure}[htbp]
\begin{center}
    \includegraphics[width=1.0\textwidth]{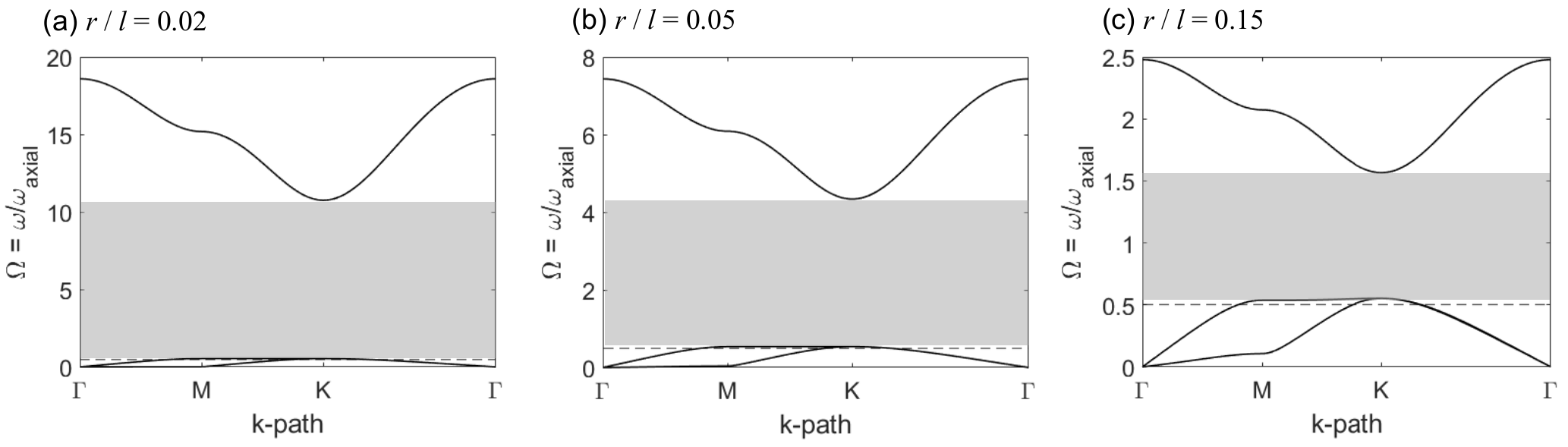}
    \caption[Effect of beam slenderness ratio on dispersion relations.]
    {Dispersion relations of the same periodic square lattice as in Fig.~\ref{fig5:4}(a) for different beam slenderness ratios: (a) $r/l = 0.02$, (b) $r/l = 0.05$, and (c) $r/l = 0.15$. Frequencies are normalized by the axial cutoff frequency $\omega_{\mathrm{axial}}$. The shaded regions indicate the first complete band gaps.  Dashed lines indicate $\Omega = 0.5$.}
    \label{fig5:5}
\end{center}
\end{figure}

We next examine the effect of beam slenderness ratio $r/l$ on the dispersion relations by varying the cross-sectional radius $r$ while keeping the beam length $l$ and material properties fixed. Here, we compute the dispersion relations for the same periodic square lattice used above, with $r/l$ varied from 0.1 to 0.02, 0.05, and 0.15, as shown in Fig.~\ref{fig5:5}(a)--(c). The results show that the onset of the first band gap remains nearly unchanged at $\Omega \approx 0.5$ across all considered $r/l$ values, indicating that the onset frequency continues to scale with $\omega_{\mathrm{axial}}$. This behavior is expected because the axial cutoff frequency $\omega_{\mathrm{axial}}$ in Eq.~\eqref{eqn:5_9} depends only on beam length and material properties and is independent of the cross-sectional radius. In contrast, the bending cutoff frequency $\omega_{\mathrm{bend}}$, which depends on cross-sectional geometry as discussed in Section~\ref{5sec:3_1}, varies with $r/l$. However, the band gap onset remains nearly unchanged over the range of $r/l$ values considered here, indicating that the onset frequency is not governed by $\omega_{\mathrm{bend}}$. At higher frequencies, the upper termination of the band gap varies significantly with $r/l$. This behavior is associated with shifts in the high-frequency rotational branches, whose frequency scale $\omega_{\mathrm{rot}}$ depends on beam geometry as discussed in Section~\ref{5sec:3_1}. As a result, variations in $r/l$ primarily affect the band gap termination, while leaving the band gap onset largely unchanged.

These results further support the argument presented in Section~\ref{5sec:4_2} that band gap onset is not governed by an individual-beam flexural resonance mechanism. If the onset frequency were controlled by the fixed-fixed flexural resonance frequency, which depends strongly on beam geometry, the band gap onset would be expected to vary significantly with beam slenderness ratio. The weak dependence of the onset frequency on $r/l$ therefore provides additional evidence against an individual-beam flexural resonance mechanism.

\section{Band Gap Onset in Disordered Beam Networks}\label{5sec:5}
Section~\ref{5sec:4} demonstrated that in periodic beam networks the first complete band gap is governed by the axial cutoff frequency $\omega_{\mathrm{axial}}$. A natural question is whether the band gap onset frequency in disordered beam networks, where connectivity and beam orientations vary spatially, is also governed by $\omega_{\mathrm{axial}}$. To address this question, we construct periodic supercells of disordered networks generated from equilibrium hard-disk configurations, random sequential addition (RSA) configurations, and binomial point process realizations. These point configurations produce distinct beam-length statistics, allowing us to examine how variations in beam-length statistics affect the band gap onset frequency.

\subsection{Periodic supercell construction of disordered beam networks}\label{5sec:5_1}
\begin{figure}[htbp]
\begin{center}
    \includegraphics[width=1.0\textwidth]{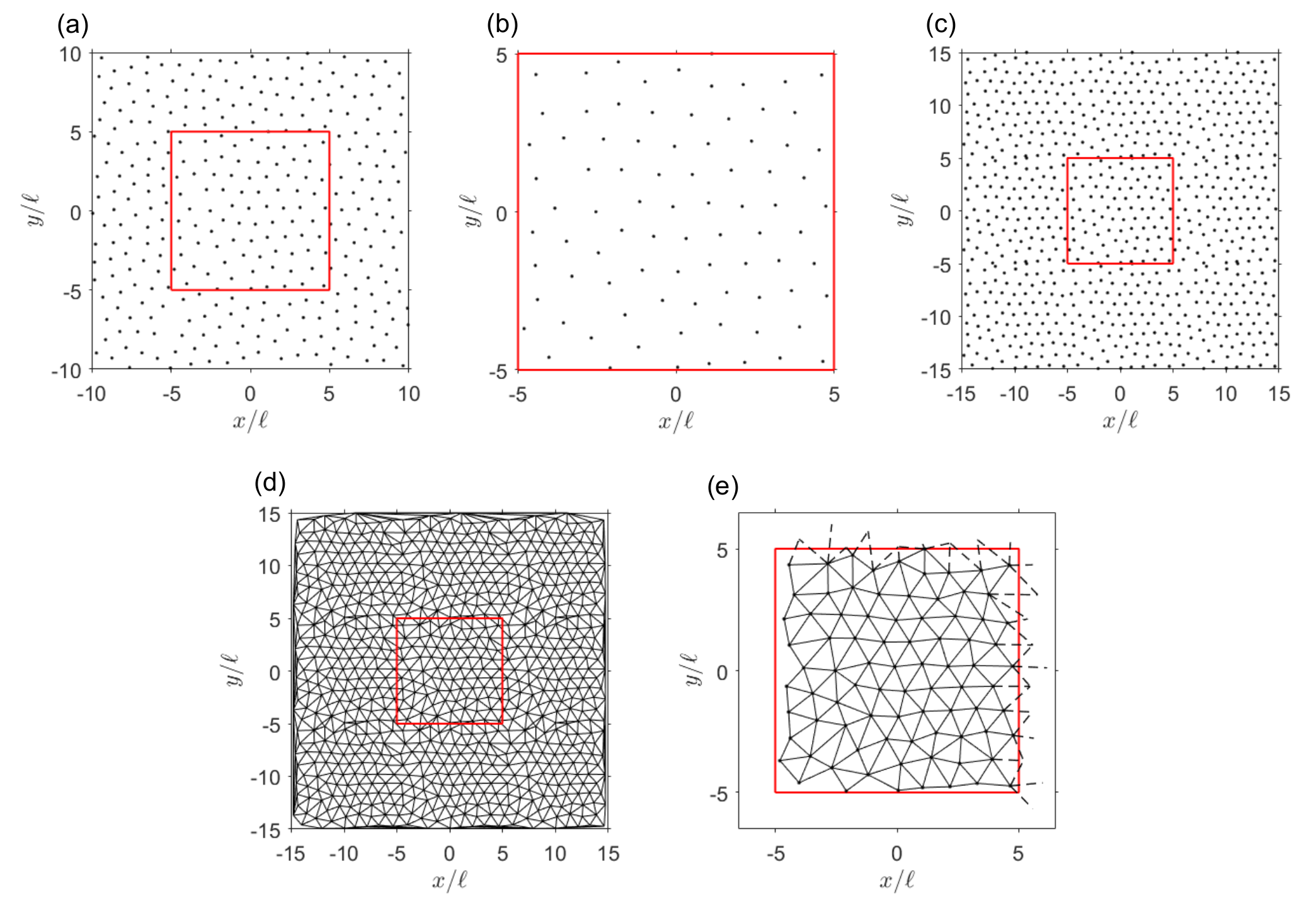}
    \caption[Periodic supercell construction for disordered beam networks.]
    {Construction of a periodic supercell of a disordered beam network (illustrated for a equilibrium hard-disk configuration). Coordinates are shown in nondimensional form $x / l$ and $y / l$, centered at the origin. (a) Initial point configuration generated in the domain $[-10, 10] \times [-10, 10]$. The red square indicates the interior region used to define the periodic supercell. (b) Points retained within the supercell, spanning $[-5, 5] \times [-5, 5]$. (c) $3 \times 3$ tiled point cloud constructed by translating the retained point set to neighboring cells. (d) Delaunay triangulation performed on the tiled point set to generate connectivity. (e) Extracted periodic supercell network. Solid edges correspond to internal connections, whereas dashed edges represent boundary-crossing periodic connections. The same construction procedure is applied to RSA configurations and binomial point process realizations.}
    \label{fig5:6}
\end{center}
\end{figure}

An infinite disordered beam network is represented using a finite periodic supercell, to which the Bloch-periodic formulation introduced in Section~\ref{5sec:2} is applied. Disordered point configurations are first generated in a square domain $[-10, 10] \times [-10, 10]$ in nondimensional coordinates $x/l$ and $y/l$ (Fig.~\ref{fig5:6}(a)), where $l$ denotes the same beam length used in the periodic networks of Sections~\ref{5sec:3} and~\ref{5sec:4}. Three representative point configurations are considered: equilibrium hard-disk configurations (strong short-range exclusion), RSA configurations (intermediate short-range exclusion), and binomial point process realizations. These point configurations generate beam networks with distinct beam-length statistics. For the equilibrium hard-disk configurations, point positions are initialized on a square lattice, with the number of points chosen to achieve an area fraction $\phi = 0.65$ based on disks of diameter $l$. The point positions are then randomized using Monte Carlo trial displacements under periodic boundary conditions, with trial moves rejected if they lead to disk overlap, thereby enforcing a minimum center-to-center separation equal to $l$. RSA configurations are generated by sequential random insertion \cite{Widom1966}, accepting only points that satisfy a minimum separation equal to $l/2$, i.e., half that used in the equilibrium hard-disk configurations. The insertion process continues until the same number of points as in the equilibrium hard-disk configurations is reached. Binomial point process realizations are generated by independently sampling the same number of points from a uniform distribution without minimum-separation constraints. To ensure that differences among the resulting networks arise primarily from the point-generation process, the same number of points and domain size are used across all configurations.

To mitigate boundary effects associated with the finite sampling window, only points located within the interior region $[-5, 5] \times [-5, 5]$ are retained (Fig.~\ref{fig5:6}(b)). This interior region, of size $10 \times 10$, defines the periodic supercell used in the subsequent eigenfrequency analysis. To ensure periodic connectivity across cell boundaries, a $3 \times 3$ tiled point cloud is constructed by translating the retained point set by integer multiples of the lattice vectors $(10, 0)$ and $(0, 10)$ (Fig.~\ref{fig5:6}(c)). Delaunay triangulation is then performed on the tiled point set to generate connectivity (Fig.~\ref{fig5:6}(d)), and edges are extracted from the triangulation. The periodic supercell network is constructed by retaining all edges that have at least one endpoint in the central tile. Edges whose two endpoints both lie outside the central tile are excluded, as they do not belong to the supercell. Edges with both endpoints inside the central tile are treated as internal connections. 

Special care is required for edges that cross the supercell boundary. In the $3 \times 3$ tiling, each node is assigned an integer lattice index $(s_x, s_y) \in \mathbb{Z}^2$, where $(0, 0)$ denotes the central supercell and nonzero indices correspond to nodes translated from the central supercell by lattice vectors of the form $\mathbf{R} = s_x (10, 0) + s_y (0, 10)$. A boundary-crossing edge connects a node in the central supercell to a node in a neighboring cell. Because lattice translation vectors that differ only by sign (i.e., $\mathbf{R}$ and $-\mathbf{R}$) represent the same periodic connection, only lattice translations satisfying $s_x > 0$, or $s_x = 0$ and $s_y > 0$, are retained. The resulting periodic supercell network is shown in Fig.~\ref{fig5:6}(e). For the retained boundary-crossing edges, Bloch periodic boundary conditions are imposed by relating the DOFs of the translated node to those of the corresponding node in the central supercell through the Bloch phase factor $e^{i\mathbf{k}\cdot\mathbf{R}}$, where $\mathbf{R} = s_x (10, 0) + s_y (0, 10)$. This treatment is consistent with the Bloch-periodic FE formulation of \cite{sukumar2009}, where DOFs separated by lattice translations are connected through Bloch phase factors. In this manner, the $10 \times 10$ supercell defines a fully periodic beam network within the Bloch-periodic FE framework introduced in Section~\ref{5sec:2}. For each point-process type, five independent realizations are generated using the above construction procedure. Each realization defines a distinct periodic supercell that is analyzed separately within the Bloch-periodic FE framework.

\subsection{Density of states and beam-length statistics}\label{5sec:5_2}
In this section, the DOS of the disordered beam networks is computed using the periodic supercells constructed in Section~\ref{5sec:5_1}. In contrast to the primitive periodic lattices considered in Section~\ref{5sec:4}, the disordered supercells do not possess sufficient lattice symmetry to justify restricting the analysis to high-symmetry paths in the Brillouin zone. Although the supercell is formally periodic, the large number of DOFs produces densely folded dispersion branches, making path-based dispersion analysis difficult to interpret. Accordingly, the DOS is evaluated over the first Brillouin zone, and the band gap onset frequency $\omega_{\mathrm{onset}}$ is defined as the lowest frequency at which the DOS first vanishes over a finite frequency interval. DOS has been used to identify band gaps in disordered photonic and phononic systems \cite{froufe2016,Gkantzounis2017}.

The same material properties and beam cross-sectional parameters used in Sections~\ref{5sec:3} and~\ref{5sec:4} are adopted here. The $10 \times 10$ periodic supercell defined in Section~\ref{5sec:5_1} is analyzed within the Bloch-periodic FE framework described in Section~\ref{5sec:2}. Because the supercell is square, the first Brillouin zone is also square, with side lengths inversely proportional to the supercell size. The Brillouin zone is uniformly sampled using a $120 \times 120$ grid in wavevector space. For each wavevector, the Bloch-reduced eigenvalue problem is solved to obtain the angular eigenfrequencies, and the resulting eigenfrequencies are used to construct the DOS. For each point-process type, five independent realizations are analyzed following this procedure. Here, frequencies are normalized as $\Omega = \omega / \omega_{\mathrm{axial}}$, where $\omega_{\mathrm{axial}}$ is the axial cutoff frequency of a 1D periodic beam of length $l$ defined in Eq.~\eqref{eqn:5_9}. This normalization facilitates direct comparison with the periodic lattices because the disordered configurations are generated using the same reference length $l$ used to define the beam length in the periodic networks.

\begin{figure}[htbp]
\begin{center}
    \includegraphics[width=1.0\textwidth]{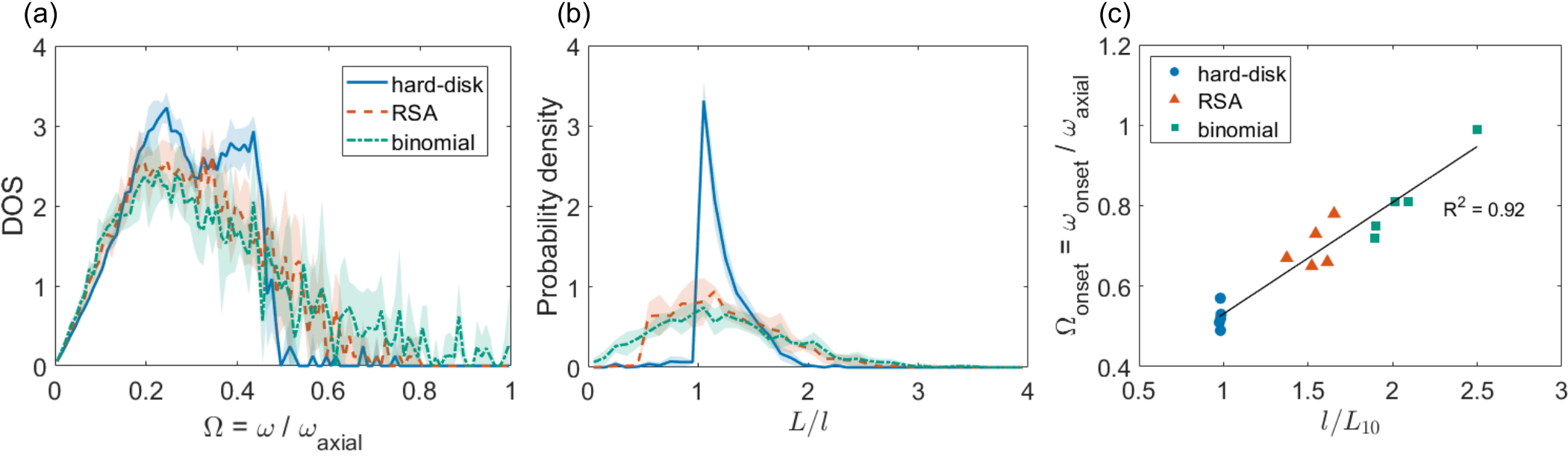}
    \caption[Density of states, beam-length statistics, and band gap onset scaling in disordered beam networks.]
    {DOS of disordered beam networks generated from equilibrium hard-disk configurations, RSA configurations, and binomial point process realizations. Frequencies are normalized by the axial cutoff frequency $\omega_{\mathrm{axial}}$. For each point-process type, five independent realizations are analyzed; solid lines represent the mean DOS and shaded regions denote $\pm 1$ standard deviation. (b) Corresponding normalized beam-length distributions $L/l$ for the same realizations, where $L$ denotes the individual beam length in the disordered networks and $l$ the beam length in the periodic networks, shown as mean curves with $\pm 1$ standard deviation shading. (c) $\Omega_{\mathrm{onset}}$ for each realization plotted against $l / L_{10}$. The solid line indicates a linear regression fit ($R^2 = 0.92$).}
    \label{fig5:7}
\end{center}
\end{figure}

Figure~\ref{fig5:7}(a) shows the mean DOS across five realizations for three disordered networks generated from equilibrium hard-disk, RSA, and binomial point processes, with shaded regions indicating $\pm 1$ standard deviation. The DOS differs systematically across the three disordered networks. For the equilibrium hard-disk network, the DOS first vanishes at $\Omega \approx 0.5$, consistent with the periodic lattices—specifically the square, triangular, and honeycomb lattices shown in Fig.~\ref{fig5:4}(a)--(c). In contrast, for the RSA and binomial networks, the DOS first vanishes at higher frequencies, $\Omega \approx 0.7$ and $0.8$, respectively. The shaded regions indicate relatively small inter-realization variability compared to the overall shift in the onset frequency among the three networks.

Fig.~\ref{fig5:7}(b) shows the corresponding normalized beam-length distributions $L/l$ for the same disordered network realizations used in Fig.~\ref{fig5:7}(a), where $L$ denotes the individual beam length in the disordered networks and $l$ the beam length in the periodic networks. The distributions are presented as mean curves with $\pm 1$ standard deviation shading. The shaded regions indicate small variability across realizations for each network. The equilibrium hard-disk networks exhibit a sharply bounded distribution with very few beams shorter than $l$, reflecting the imposed minimum separation constraint (minor deviations arise due to the periodic tiling step illustrated in Fig.~\ref{fig5:6}(c), where the minimum-separation constraint is not strictly enforced between translated points near the boundaries). The RSA networks show a broader distribution with a noticeable fraction of beams shorter than $l$, reflecting the reduced minimum-separation constraint ($l/2$), whereas the binomial networks exhibit the broadest distribution, including a substantial short-beam tail due to the absence of a minimum-separation constraint. These systematic differences in the short-beam tail of the distribution suggest that variations in beam-length statistics may influence the location of the band gap onset.

Section~\ref{5sec:4} showed that the onset of the first complete band gap in periodic lattices scales with $\omega_{\mathrm{axial}}$. Since the analytical dispersion relation of the 1D periodic beam in Section~\ref{5sec:3_1} gives $\omega_{\mathrm{axial}} \propto 1/l$ (Eq.~\eqref{eqn:5_9}), the band gap onset frequency scales inversely with beam length. Extending this argument to disordered networks suggests that $\omega_{\mathrm{onset}}$ should correlate with the inverse of a characteristic short-beam length. For each realization, the normalized band gap onset frequency $\Omega_{\mathrm{onset}} = \omega_{\mathrm{onset}} / \omega_{\mathrm{axial}}$ is identified as the lowest normalized frequency at which the DOS first vanishes over a finite interval of width $\Delta \Omega \ge 0.05$, thereby avoiding spurious isolated zero-density bins arising from finite Brillouin-zone sampling. We adopt the 10th percentile beam length, $L_{10}$, as a characteristic short-beam length. Figure~\ref{fig5:7}(c) shows $\Omega_{\mathrm{onset}}$ for each realization against $l / L_{10}$. A clear approximately linear trend with a positive slope is observed across realizations from all network types, with $R^2 = 0.92$, indicating that the band gap onset frequency increases with the inverse of the characteristic short-beam length. These results quantitatively support the scaling of band gap onset with the axial cutoff frequency identified for the periodic lattices and indicate that short-beam length statistics play a governing role in band gap onset in disordered beam networks. Similar correlations are obtained when alternative values of $\Delta \Omega$ (e.g., 0.03 or 0.07) are used to identify band gap onset, indicating that the observed correlation is not sensitive to the specific choice of $\Delta \Omega$. Likewise, similar correlations are obtained when alternative percentiles (e.g., 5th or 15th) are used to define the characteristic short-beam length, indicating that the observed correlation is not sensitive to the specific percentile choice.

\section{Conclusions}\label{5sec:6}
This study shows that band gap formation in beam networks is governed by geometry-induced coupling between deformation modes. Band gap onset arises from axial–bending coupling, causing axial stiffness to dominate the effective network response. As a result, band gap onset is governed by the axial cutoff frequency. The band gap termination is governed by higher-frequency rotational dynamics. These results indicate that conventional Bragg scattering and local resonance mechanisms do not govern band gap formation in beam networks. The same underlying mechanism extends to both periodic and disordered beam networks. In both systems, band gap onset is governed by the axial cutoff frequency. In periodic lattices, this dependence manifests through beam orientations at lattice nodes, whereas in disordered networks it manifests through short-beam statistics arising from variations in beam length.

While the adopted hybrid mass matrix reproduces the low-frequency dispersion behavior with sufficient accuracy to identify band gap onset, the rotational branch still exhibits quantitative deviation from the analytical dispersion relation of the 1D periodic beam in the high-frequency regime. Although this deviation does not affect the band gap onset frequency, it may influence the quantitative prediction of the band gap termination frequency and, consequently, the band gap width. Improved modeling of rotational dynamics may therefore enable more accurate prediction of the high-frequency dispersion behavior and the resulting band gap termination frequency. 

In addition, the present analysis is restricted to 2D in-plane beam networks and does not account for out-of-plane deformation or torsional modes. Extending the framework to fully three-dimensional beam lattices would clarify how additional deformation modes influence band gap formation in beam networks.

\section*{Data Availability}
The MATLAB codes used to generate the results presented in this study are available at \url{https://github.com/DMREF-networks/beam-network-bandgap-mechanism}.

\section*{Acknowledgements}
This research was supported by the U.S. National Science Foundation through DMREF Grants No. CMMI-2323344 and No. CMMI-2323342. The authors thank Karen E. Daniels for helpful comments and feedback on the manuscript.

\appendix
\section{Comparison of Finite Element and Analytical Dispersion Relations for Conventional Mass Matrices} \label{sec:appendix-D}
\setcounter{figure}{0}
\renewcommand{\thefigure}{A\arabic{figure}}

\setcounter{table}{0}
\renewcommand{\thetable}{A\arabic{table}}
Section~\ref{5sec:3_1} derives the analytical dispersion relations of the 1D Bloch-periodic Timoshenko beam, shown in Fig.~\ref{fig5:2}(a). This appendix compares those analytical results with FE dispersion relations obtained using conventional consistent and lumped mass matrices. The hybrid mass matrix adopted in Section~\ref{5sec:3_3} is motivated by discrepancies between the analytical dispersion relations and the FE dispersion relations, which are quantified here.

For the axial component, the conventional consistent and lumped mass blocks correspond to the consistent and lumped components appearing in Eq.~\eqref{eqn:5_17}. For the transverse and rotational components, the lumped mass block corresponds to Eq.~\eqref{eqn:5_18}, while the consistent mass block corresponds to the classical Timoshenko-beam formulation \cite{przemieniecki1985}, which accounts for both shear deformation and rotary inertia.

\begin{figure}[htbp]
\begin{center}
    \includegraphics[width=0.9\textwidth]{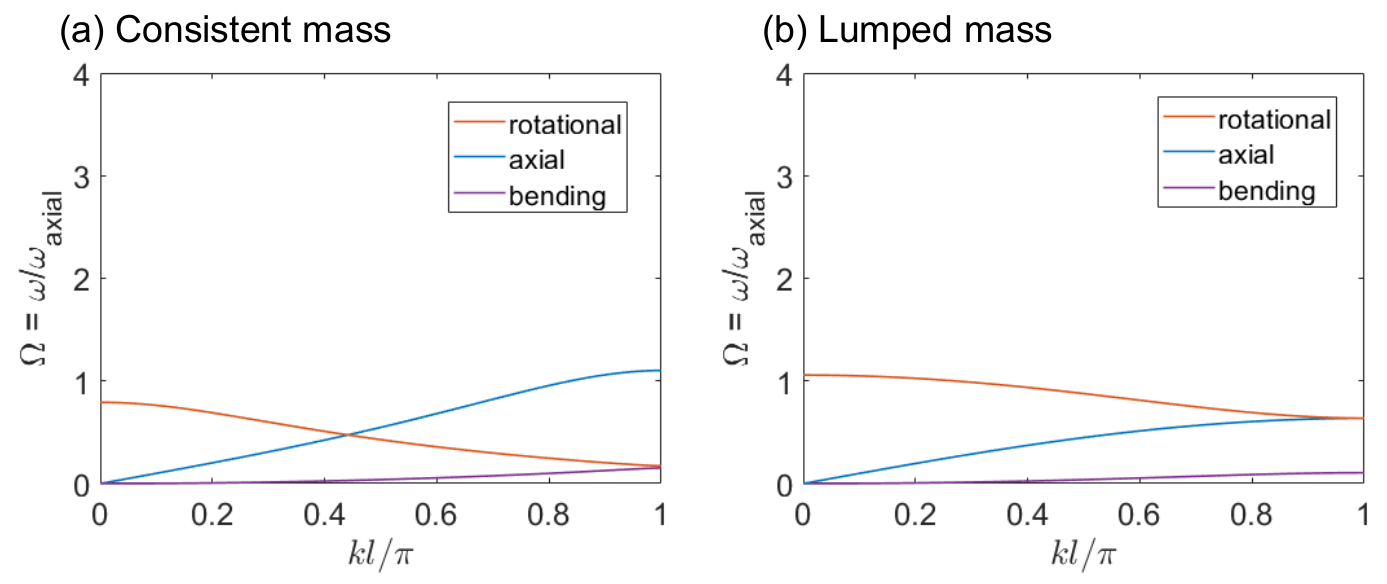}
    \caption[Dispersion relations from finite element models with consistent and lumped mass matrices.]
    {Dispersion relations of a 1D Bloch-periodic Timoshenko beam obtained from the FE models using (a) a consistent mass matrix and (b) a lumped mass matrix. Frequencies are normalized by the axial cutoff frequency $\omega_{\mathrm{axial}}$.}
    \label{figD:1}
\end{center}
\end{figure}

Figure~\ref{figD:1} shows the FE dispersion relations obtained using conventional consistent and lumped mass matrices. The material properties and beam geometry are the same as those given in Section~\ref{5sec:3_4}. Frequencies are normalized by the axial cutoff frequency $\omega_{\mathrm{axial}}$. In Fig.~\ref{figD:1}(a), the consistent mass matrix reproduces the bending branch with good qualitative agreement relative to the analytical dispersion relations shown in Fig.~\ref{fig5:2}(a), while the axial branch is moderately overestimated. By contrast, the rotational branch is underestimated relative to the analytical dispersion relations and drops below the axial branch over part of the Brillouin zone. This incorrect branch ordering places the rotational branch at lower frequencies than in the analytical dispersion relations shown in Fig.~\ref{fig5:2}(a), which can lead to incorrect identification of both band gap onset and termination in 2D beam networks. In Fig.~\ref{figD:1}(b), the lumped mass matrix systematically underestimates the axial, bending, and rotational branches relative to the analytical dispersion relations shown in Fig.~\ref{fig5:2}(a), although the overall dispersion trends remain qualitatively similar. The branch ordering is largely preserved, with the rotational branch remaining above the axial branch. The main consequence is therefore a systematic underestimation of the axial and bending frequency scales, leading to underprediction of band gap onset frequencies. Underestimation of the rotational branch also leads to underprediction of band gap termination frequencies.

These results demonstrate that neither the conventional consistent nor the lumped mass matrix fully reproduces the analytical dispersion relations of the 1D Bloch-periodic Timoshenko beam. The hybrid mass matrix introduced in Section~\ref{5sec:3_3} is therefore adopted to improve agreement with the analytical dispersion relations in both branch ordering and frequency values.


\bibliographystyle{elsarticle-num}
\bibliography{references}

@book{chandrupatla2012fem,
  author = {Chandrupatla, T. R. and Belegundu, A. D.},
  title = {Introduction to Finite Elements in Engineering},
  edition = {4},
  publisher = {Pearson},
  year = {2012}
}

@article{askes2024,
  author = {Askes, H. and Lombardo, M. and Nguyen, D. C.},
  title = {Homogenisation of periodic lattices with lumped and distributed mass: Beam models, continualisation and stabilisation},
  journal = {International Journal of Solids and Structures},
  volume = {302},
  pages = {112988},
  year = {2024}
}

@article{chen2017,
  author = {Chen, Y. and Qian, F. and Zuo, L. and Scarpa, F. and Wang, L.},
  title = {Broadband and multiband vibration mitigation in lattice metamaterials with sinusoidally-shaped ligaments},
  journal = {Extreme Mechanics Letters},
  volume = {17},
  pages = {24--32},
  year = {2017}
}

@article{cowper1966,
  author = {Cowper, G.},
  title = {The shear coefficient in Timoshenko’s beam theory},
  journal = {Journal of Applied Mechanics},
  volume = {33},
  number = {2},
  pages = {335--340},
  year = {1966}
}

@article{froufe2016,
  author = {Froufe-P{\'e}rez, L. S. and Engel, M. and Damasceno, P. F. and Muller, N. and Haberko, J. and Glotzer, S. C. and Scheffold, F.},
  title = {Role of short-range order and hyperuniformity in the formation of band gaps in disordered photonic materials},
  journal = {Physical Review Letters},
  volume = {117},
  number = {5},
  pages = {053902},
  year = {2016}
}

@article{hussein2014,
  author = {Hussein, M. I. and Leamy, M. J. and Ruzzene, M.},
  title = {Dynamics of phononic materials and structures: Historical origins, recent progress, and future outlook},
  journal = {Applied Mechanics Reviews},
  volume = {66},
  number = {4},
  pages = {040802},
  year = {2014}
}

@article{joannopoulos1997,
  author = {Joannopoulos, J. D. and Villeneuve, P. R. and Fan, S.},
  title = {Photonic crystals: Putting a new twist on light},
  journal = {Nature},
  volume = {386},
  number = {6621},
  pages = {143--149},
  year = {1997}
}

@book{joannopoulos2008,
  author = {Joannopoulos, J. D. and Johnson, S. G. and Winn, J. N. and Meade, R. D.},
  title = {Photonic Crystals: Molding the Flow of Light},
  publisher = {Princeton University Press},
  address = {Princeton, NJ},
  year = {2008}
}

@article{kim1993,
  author = {Kim, K. O.},
  title = {A review of mass matrices for eigenproblems},
  journal = {Computers \& Structures},
  volume = {46},
  number = {6},
  pages = {1041--1048},
  year = {1993}
}

@article{laier2007,
  author = {Laier, J. E.},
  title = {Mass lumping, dispersive properties and bifurcation of Timoshenko’s flexural waves},
  journal = {Advances in Engineering Software},
  volume = {38},
  number = {8--9},
  pages = {547--551},
  year = {2007}
}

@book{laude2015,
  author = {Laude, V.},
  title = {Phononic Crystals: Artificial Crystals for Sonic, Acoustic, and Elastic Waves},
  publisher = {Walter de Gruyter},
  year = {2015}
}

@article{lu2009,
  author = {Lu, M. H. and Feng, L. and Chen, Y. F.},
  title = {Phononic crystals and acoustic metamaterials},
  journal = {Materials Today},
  volume = {12},
  number = {12},
  pages = {34--42},
  year = {2009}
}

@article{maldovan2013,
  author = {Maldovan, M.},
  title = {Sound and heat revolutions in phononics},
  journal = {Nature},
  volume = {503},
  number = {7475},
  pages = {209--217},
  year = {2013}
}

@article{phani2006,
  author = {Phani, A. S. and Woodhouse, J. and Fleck, N. A.},
  title = {Wave propagation in two-dimensional periodic lattices},
  journal = {The Journal of the Acoustical Society of America},
  volume = {119},
  number = {4},
  pages = {1995--2005},
  year = {2006}
}

@book{przemieniecki1985,
  author = {Przemieniecki, J. S.},
  title = {Theory of Matrix Structural Analysis},
  publisher = {Courier Corporation},
  year = {1985}
}

@book{rao2017,
  author = {Rao, S. S.},
  title = {Mechanical Vibrations, SI Units},
  publisher = {Pearson},
  address = {London},
  year = {2017}
}

@article{sigmund2003,
  author = {Sigmund, O. and Jensen, J. S.},
  title = {Systematic design of phononic band-gap materials and structures by topology optimization},
  journal = {Philosophical Transactions of the Royal Society A},
  volume = {361},
  number = {1806},
  pages = {1001--1019},
  year = {2003}
}

@article{sukumar2009,
  author = {Sukumar, N. and Pask, J. E.},
  title = {Classical and enriched finite element formulations for Bloch-periodic boundary conditions},
  journal = {International Journal for Numerical Methods in Engineering},
  volume = {77},
  number = {8},
  pages = {1121--1138},
  year = {2009}
}

@article{trainiti2016,
  author = {Trainiti, G. and Rimoli, J. J. and Ruzzene, M.},
  title = {Wave propagation in undulated structural lattices},
  journal = {International Journal of Solids and Structures},
  volume = {97},
  pages = {431--444},
  year = {2016}
}

@article{wang2015,
  author = {Wang, P. and Casadei, F. and Kang, S. H. and Bertoldi, K.},
  title = {Locally resonant band gaps in periodic beam lattices by tuning connectivity},
  journal = {Physical Review B},
  volume = {91},
  number = {2},
  pages = {020103},
  year = {2015}
}

@article{warmuth2015,
  author = {Warmuth, F. and K{\"o}rner, C.},
  title = {Phononic band gaps in 2D quadratic and 3D cubic cellular structures},
  journal = {Materials},
  volume = {8},
  number = {12},
  pages = {8327--8337},
  year = {2015}
}

@article{Svensson2010,
  author  = {Svensson, Jonas and Andersson, Patrik and Kropp, Wolfgang},
  title   = {On the design of structural junctions for the purpose of hybrid passive-active vibration control},
  journal = {J. Sound Vib.},
  volume  = {329},
  pages   = {1274--1288},
  year    = {2010}
}

@book{Cremer2005,
  author    = {Cremer, L. and Heckl, M. and Petersson, B. A. T.},
  title     = {Structure-borne sound: structural vibrations and sound radiation at audio frequencies},
  publisher = {Springer},
  year      = {2005}
}

@article{Leamy2012,
  author  = {Leamy, Michael J.},
  title   = {Exact wave-based Bloch analysis procedure for investigating wave propagation in two-dimensional periodic lattices},
  journal = {J. Sound Vib.},
  volume  = {331},
  number  = {7},
  pages   = {1580--1596},
  year    = {2012},
}

@article{Liu2002,
  title = {Three-component elastic wave band-gap material},
  author = {Liu, Zhengyou and Chan, C. T. and Sheng, Ping},
  journal = {Phys. Rev. B},
  volume = {65},
  issue = {16},
  pages = {165116},
  numpages = {6},
  year = {2002},
  publisher = {American Physical Society},
}

@article{Liu2012,
    author = {Liu, Liao and Hussein, Mahmoud I.},
    title = {Wave Motion in Periodic Flexural Beams and Characterization of the Transition Between Bragg Scattering and Local Resonance},
    journal = {Journal of Applied Mechanics},
    volume = {79},
    number = {1},
    pages = {011003},
    year = {2012},
    issn = {0021-8936},
}

@article{Liu2000,
author = {Zhengyou Liu  and Xixiang Zhang  and Yiwei Mao  and Y. Y. Zhu  and Zhiyu Yang  and C. T. Chan  and Ping Sheng },
title = {Locally Resonant Sonic Materials},
journal = {Science},
volume = {289},
number = {5485},
pages = {1734-1736},
year = {2000},
}

@article{Gkantzounis2017,
  title = {Hyperuniform disordered phononic structures},
  author = {Gkantzounis, G. and Amoah, T. and Florescu, M.},
  journal = {Phys. Rev. B},
  volume = {95},
  issue = {9},
  pages = {094120},
  numpages = {11},
  year = {2017},
  publisher = {American Physical Society},
}

@article{maldovan2015,
  title={Phonon wave interference and thermal bandgap materials},
  author={Maldovan, Martin},
  journal={Nature materials},
  volume={14},
  number={7},
  pages={667--674},
  year={2015},
  publisher={Nature Publishing Group UK London}
}

@article{Widom1966,
    author = {Widom, B.},
    title = {Random Sequential Addition of Hard Spheres to a Volume},
    journal = {The Journal of Chemical Physics},
    volume = {44},
    number = {10},
    pages = {3888-3894},
    year = {1966},
    issn = {0021-9606},
}

@article{klatt2022,
  title={Wave propagation and band tails of two-dimensional disordered systems in the thermodynamic limit},
  author={Klatt, Michael A and Steinhardt, Paul J and Torquato, Salvatore},
  journal={Proceedings of the National Academy of Sciences},
  volume={119},
  number={52},
  pages={e2213633119},
  year={2022},
  publisher={National Academy of Sciences}
}

@article{bertoldi2017,
  title={Flexible mechanical metamaterials},
  author={Bertoldi, Katia and Vitelli, Vincenzo and Christensen, Johan and Van Hecke, Martin},
  journal={Nature Reviews Materials},
  volume={2},
  number={11},
  pages={1--11},
  year={2017},
  publisher={Nature Publishing Group}
}

@article{craster2023,
  title={Mechanical metamaterials},
  author={Craster, Richard and Guenneau, S{\'e}bastien and Kadic, Muamer and Wegener, Martin},
  journal={Reports on Progress in Physics},
  volume={86},
  number={9},
  pages={094501},
  year={2023},
  publisher={IOP Publishing}
}

@article{jiao2023,
  title={Mechanical metamaterials and beyond},
  author={Jiao, Pengcheng and Mueller, Jochen and Raney, Jordan R and Zheng, Xiaoyu and Alavi, Amir H},
  journal={Nature communications},
  volume={14},
  number={1},
  pages={6004},
  year={2023},
  publisher={Nature Publishing Group UK London}
}

@misc{maher2026,
      title={Probing the influence of topological and geometric disorder on the spectrum of the differential Laplacian operator on networks}, 
      author={Charles Emmett Maher and Jeremy L. Marzuola and Katherine A. Newhall},
      year={2026},
      eprint={2602.22341},
      archivePrefix={arXiv},
      primaryClass={cond-mat.dis-nn},
      url={https://arxiv.org/abs/2602.22341}, 
}

@misc{moody2026,
      title={A Methodology for Manufacturing 3D Disordered Metamaterials Using Laser Powder Bed Fusion from Granular Packings and Hyperuniform Point Clouds}, 
      author={Katherine Moody and Molly Li and Charles Emmett Maher and Kwangmin Lee and Timothy Horn and Katherine A. Newhall and Ryan Hurley and Karen E. Daniels and Christopher Rock},
      year={2026},
      eprint={5865},
      archivePrefix={engrXiv},
      url={https://doi.org/10.31224/5865}, 
}

@article{siedentop_stealthy_2024,
	title = {Stealthy and hyperuniform isotropic photonic band gap structure in {3D}},
	volume = {3},
	copyright = {https://creativecommons.org/licenses/by/4.0/},
	issn = {2752-6542},
	number = {9},
	journal = {PNAS Nexus},
	author = {Siedentop, Lukas and Lui, Gianluc and Maret, Georg and Chaikin, Paul M and Steinhardt, Paul J and Torquato, Salvatore and Keim, Peter and Florescu, Marian},
	editor = {Amon, Cristina H},
	month = sep,
	year = {2024},
	pages = {pgae383},
}

@article{torquato_multifunctional_2018,
	title = {Multifunctional hyperuniform cellular networks: optimality, anisotropy and disorder},
	volume = {1},
	issn = {2399-7532},
	shorttitle = {Multifunctional hyperuniform cellular networks},
	number = {1},
	journal = {Multifunct. Mater.},
	author = {Torquato, S and Chen, D},
	year = {2018},
	pages = {015001},
}

@article{reid_auxetic_2018,
	title = {Auxetic metamaterials from disordered networks},
	volume = {115},
	number = {7},
	journal = {Proc. Natl. Acad. Sci.},
	author = {Reid, Daniel R. and Pashine, Nidhi and Wozniak, Justin M. and Jaeger, Heinrich M. and Liu, Andrea J. and Nagel, Sidney R. and de Pablo, Juan J.},
	year = {2018},
	pages = {E1384--E1390},
}

@article{shen_autonomous_2024,
	title = {An autonomous design algorithm to experimentally realize three-dimensionally isotropic auxetic network structures without compromising density},
	volume = {10},
	copyright = {2024 The Author(s)},
	issn = {2057-3960},
	number = {1},
	journal = {npj Comput. Mater.},
	author = {Shen, Meng and Reyes-Martinez, Marcos A. and Powell, Louise Ahure and Iadicola, Mark A. and Sharma, Abhishek and Byléhn, Fabian and Pashine, Nidhi and Chan, Edwin P. and Soles, Christopher L. and Jaeger, Heinrich M. and de Pablo, Juan J.},
	year = {2024},
	pages = {113},
}

@article{rreid_ideal_2019,
  title={Ideal isotropic auxetic networks from random networks},
  author={Reid, Daniel R and Pashine, Nidhi and Bowen, Alec S and Nagel, Sidney R and de Pablo, Juan J},
  journal={Soft Matter},
  volume={15},
  number={40},
  pages={8084--8091},
  year={2019},
  publisher={Royal Society of Chemistry},
}

@article{areyes-martinez_tuning_2022,
	title = {Tuning the mechanical impedance of disordered networks for impact mitigation},
	volume = {18},
	number = {10},
	journal = {Soft Matter},
	author = {Reyes-Martinez, Marcos A. and Chan, Edwin P. and Soles, Christopher L. and Han, Endao and Murphy, Kieran A. and Jaeger, Heinrich M. and Reid, Daniel R. and Pablo, Juan J. de},
	year = {2022},
	pages = {2039--2045},
}

@article{mendels_systematic_2023,
	title = {Systematic modification of functionality in disordered elastic networks through free energy surface tailoring},
	volume = {9},
	number = {23},
	journal = {Sci. Adv.},
	author = {Mendels, Dan and Byléhn, Fabian and Sirk, Timothy W. and de Pablo, Juan J.},
	month = jun,
	year = {2023},
	pages = {eadf7541},
}

@article{man_photonic_2013,
	title = {Photonic band gap in isotropic hyperuniform disordered solids with low dielectric contrast},
	volume = {21},
	number = {17},
	urldate = {2021-08-11},
	journal = {Optics Express},
	author = {Man, Weining and Florescu, Marian and Matsuyama, Kazue and Yadak, Polin and Nahal, Geev and Hashemizad, Seyed and Williamson, Eric and Steinhardt, Paul and Torquato, Salvatore and Chaikin, Paul},
	year = {2013},
	pages = {19972--19981},
}

@article{florescu_designer_2009,
	title = {Designer disordered materials with large, complete photonic band gaps},
	volume = {106},
	issn = {0027-8424, 1091-6490},
	number = {49},
	journal = {Proc. Natl. Acad. Sci.},
	author = {Florescu, Marian and Torquato, Salvatore and Steinhardt, Paul J.},
	year = {2009},
	pages = {20658--20663},
}


\end{document}